# Comments to "High current ionic flows via ultra-fast lasers for fusion applications" by H. Ruhl and G. Korn (Marvel Fusion, Munich) [1]

K. Lackner, S. Fietz, A. v. Müller

Max Planck Institute for Plasma Physics, 85748 Garching, Germany

Abstract

We comment on Marvel Fusion's proposal to use ultra-short, intensive laser pulses impinging on nanostructured targets for the production of hot ions, with the ultimate aim of energy production through the p-$^{11}$B fusion reaction. We show how a minimum requirement on the line density of the heated region arises also for such schemes aiming at ignition or just beam-energy amplification, which already for the D-T reaction are prohibitive if no pre-compression of the target is foreseen. We conclude that advanced ultra-short laser pulses can be of use for fusion applications only if the fusion reaction time can be brought down to sufficiently short values. This can happen only via strong pre-compression of the target, which appears difficult to reconcile with a nanostructured target and in particular with the fuel assembly geometries shown by the proponents.

keywords: aneutronic fusion, laser fusion, uncompressed targets, magneto-inertial confinement, beam fusion, nanostructured fusion targets, proton-boron reaction

## 1. Introduction

The ultimate goal of Marvel Fusion is fusion energy production using aneutronic fuels like p-$^{11}$B [2]. Their proposed schemes and geometries rely on the efficient production of high-energy ions by femtosecond pulse lasers in combination with nanostructured targets to improve the laser-plasma coupling. The integrity of the nanostructured targets up to the arrival of the laser pulses, the quoted plasma parameters, and the proposed geometries imply that targets are at solid-state or lower densities[i]. Their most recent paper [1] describes in somewhat more detail the development and the results of a code analyzing the production of hot ions by a laser pulse, and in the planned ultimate development stage also the thermonuclear burn.

We show in this comment, that the use of fast ions in different target configurations could not alleviate the need for plasma pre-compression by some 3 orders of magnitude in density, and how this requirement will - at the end - also emanate from results of the code of the proponents. We do not address the laser-target interaction aspects of their proposal in detail, restricting the comments to our field of competence, fusion physics.

## 2. The universal need for high target-compression for inertial-confinement based fusion energy production

It is reported in standard inertial-fusion textbooks [3] that thermonuclear ignition requires a minimum value of the product $\varrho R$ of the region heated to burn temperature ($\varrho$ is plasma mass density and $R$ a radius). The fuel mass associated with this minimum line density and the total energy to be produced by thermonuclear reactions increases therefore strongly with *decreasing* density ($\sim 1/\varrho^2$). For D-T fuel at solid state density, this fusion energy would correspond to a small thermonuclear device[ii]. Using p-$^{11}$B as fuel would even lead to much larger unit sizes. Therefore, laser driven inertial fusion proposals based on self-consistent calculations universally assume a fuel compression by a factor on the order of 1000 in density [3].

---
[i] The nanostructure of the targets implies an averaged density below solid state.

Although pre-compression of the fuel is now parenthetically mentioned in Ref. [1], possibly following the strong criticism brought forward in Ref. [4], the target geometries described, and the parameters quoted are still based on an uncompressed fuel and a nanostructured laser-fuel interaction zone.

The justification for their dissenting opinion evidently originates from the fact that modern laser systems can concentrate energy in time and space in a previously un-envisaged manner and could thus trigger thermonuclear burn. This view ignores the fact that *thermonuclear fusion at solid state densities* is a much too slow process to profit from the short time scale of this energy concentration. Thermonuclear burn is a competition between thermonuclear reactions and thermal expansion of the target. Particle density enhances burn and - for a given energy density - retards the thermal disintegration. The balance of these two effects - at given plasma temperature - is expressed by the $\varrho R$-criterion. It is energetically expedient that a compression should proceed as much as possible during a cold stage of the matter, with ignition temperatures reached just at the final stage. This is the basis of all inertial fusion hot-spot ignition scenarios and - even more optimized - "fast ignition"-concepts.

The Marvel team, however, continues to pursue the possibility of thermonuclear energy production without pre-compression of the target. In the present paper an ambitious code development is sketched, which should ultimately support these claims[iii]. Such codes are certainly necessary for detailed experimental design, and also for estimates when effects of comparable strength compete. However, they are not necessary when competing effects differ in their intrinsic time-scales by several orders of magnitude. There exist some *necessary* criteria that any fusion concept must meet in order to produce a positive energy gain, which can be robustly evaluated in simple estimates.

The requirement of a minimum value of $\varrho R$, in particular, should appear in any proposed fusion scheme not relying on externally applied magnetic fields[iv], and hence ultimately also in the calculations with the Marvel code package once they include a sufficiently comprehensive physics model for the burn phase. As examples how this should happen, we highlight this for the sample configurations and fusion schemes proposed in [1].

At least as initial stage, the schemes suggested by Marvel Fusion rely on an energy amplification by fusion reactions of fast ions produced by laser-plasma interaction with a background fuel. This has to happen before the fast ions lose their energy (mainly due to drag on electrons). This gives rise to a necessary requirement which can be formulated by comparing the slowing-down length due to friction with the distance over which an ion has to travel to produce on average fusion reactions with an energy yield corresponding to its starting energy.

---

[ii] An equimolar D-T target at density of 0.225 g/cm$^3$, using equation 2.27 from Ref [3] for burn-up, with burn parameter H$_B$ = 7 g/cm$^2$ and a burn-up fraction of 12.5% gives an energy production of 3.4x10$^{12}$ J, equivalent to 0.85 kt TNT.

[iii] The code is only sketched in Ref. [1] and results described there refer only to the first phase: the acceleration of ions by the laser irradiation.

[iv] Or gravity, if stellar dimensions are considered; self-generated magnetic fields alone, as proposed in an earlier paper of the proponents [2]. cannot produce confinement, as we have shown in our previous comments [4], based on the virial theorem.

This is inherent also in the calculations of the proponents and in the concept of the conversion fractions $\eta^{DT}$ and $\eta^{pB}$ in Ref. [2]. In fact, their results for this *necessary criterion* are in qualitative agreement with the more elaborate calculations in Refs. [5,6]: in the case of D-T, for sufficiently high electron temperatures ($\gtrsim 5 keV$) it can be readily satisfied. For the p-$^{11}$B reaction, however, even the proponents [2] find that electron drag will stop the protons for a boron line density $n_B R_p \approx 10^{26} m^{-2}$, yielding a conversion factor $\eta^{pB} \lesssim 0.012$, and therefore an average fusion production per fast proton-ion of only 100 keV - for a proton energy of 600 keV! Even if more refined analysis does not completely rule out the possibility of breakeven for p-$^{11}$B in the absence of non-radiative energy losses [7], the required line densities in laser fusion schemes would be about one order of magnitude larger at 6 × larger ion temperatures!

Considering further necessary criteria we thus limit ourselves to the most feasible D-T reaction. The above slowing-down criterion is not sufficient if the fast ion leaves the fuel pellet before causing fusion reactions. Therefore, in a pellet with limited size, the real constraint is not slowing down, but finite dimension. To initiate thermonuclear burn, the fusion reactions (and slowing down of the reaction products) must occur predominantly within the heated fuel pellet - the "hot spot" - in order to realize ignition. This results in a minimum $R$ not given by the slowing-down distance, but by the hot-spot size of the fuel pellet[v]. In fact, the tritium line-density quoted by Marvel scientists [2] as necessary to produce a 5% conversion fraction can be taken as this criterion, but with $R_D$ to be interpreted not as slowing-down, but as hot spot radius. At optimum ion energy ($\approx 100\ keV$) it corresponds to a mass-line density of $\varrho_h R_h = 0.04$ g/cm$^2$ ($h$ for hot spot). This is significantly lower than canonical values for ignition in D-T, which refer, however, to lower energies and a thermal plasma, with $T_e \approx T_i$. In fact, repeating this estimate with 5 keV, substituting the cross-section by $\langle \sigma v \rangle / v_{th}$, yields a factor of 2 below typical simulation results[vi] for the needed hot-spot, as this is still an optimistic estimate insufficient to ensure continuing burn. Nevertheless, the value for the hot-ion regime of Marvel Fusion is far beyond any conceivable hot-spot size, in particular if densities are substituted by average densities in the nano-rod region, which might be in the range of critical densities at the appropriate laser frequency [2]. As a measure of the penalty for the lack of pre-compression we can estimate the ratio of the hot-spot energies needed at this start-of-burn point, given by $\sim \varrho_h R_h^3 \langle E \rangle \sim \frac{(\varrho_h R_h)^3}{\varrho_h^2} \langle E \rangle$, where $\langle E \rangle$ is the energy associated with the fusing ion pair (including, for a thermal case, that of the electrons). Even benefitting from a lower $\varrho_h R_h$ and taking solid state (not the critical) density for Marvel fusion, renouncing the typically 1000-fold compression implies 5 orders of magnitude increase in hot-spot energy.

## 3. Non-ignited ("driven") scenarios

Self-heating ("ignition") of a plasma is not necessary for a positive energy balance, if the fusion yield per fast ion can exceed also all energy losses associated with its creation and acceleration. This has been widely discussed and analyzed in the magnetic confinement fusion community - particularly in the context of mirror-based devices - where magnetic fields can prevent the ion escape from the fuel region. To produce this energy, the fusion reactions would not have to be confined to the hot spot region, and hence only the slowing-down and not the confinement criterion would need to be satisfied. Slowing-down, however, strongly depends on electron temperature, and the numerical value of resistivity divided by density implied by the authors in Ref. [2] for their range-estimates corresponds to $T_e \approx 4.5\ keV$, which is also generally recognized as lower limit for D on T "beam-target" fusion in magnetically confined plasmas

---

[v] An alternative justification of this estimate is that a hot ion should not carry its kinetic energy out of the hot spot region, before having left an equal amount of fusion energy behind.
[vi] see pg. 59, Ref. [3]

[5]. For lower temperatures, friction on electrons will slow down the deuterium beam too fast for fusion reactions to produce a net energy gain.

For a scheme in which ions are produced in an accelerator region, and give rise to fusion in a target region (which could be the "convertor" in Fig. 5 of Ref. [1], but also simply the region surrounding every nanostructure) the electrons in the latter still would have to be heated to typically this temperature $T_e \gtrsim 4.5\ keV$ over a region with sufficient line density for fusion energy production to exceed the invested energy. This reduces the energy requirements given above by a factor $\sim \frac{3}{2}kT_e/E_D$, (with $E_D$ the energy of the hot deuterium ions), but they still scale with $\sim 1/\varrho^2$, and would hence be exorbitant for solid state densities. Like in the ignition case, a region with a given line density has to be heated to a certain temperature, and this becomes impossible if the energy must be distributed among too many particles. The number of particles corresponding to a given line density $\varrho R$, however, is $\sim \varrho R^3 \sim (\varrho R)^3/\varrho^2$, and thus - for constant $\varrho R$ - increases quadratically with decreasing density.

To derive the $\varrho R$-criterion and its consequences we have so far treated a spherical geometry and ion motion as isotropic. For the ignition, this can be justified for these order-of magnitude comparisons as already the first-generation alpha-particles will be isotropic. Thermonuclear burn - as competition of a volume gain and a surface loss energy - will be optimized by such a spherical geometry. For the beam-driven energy amplification scenarios discussed above, ions rigorously propagating only in a plane, as suggested in Ref. [1], would be experiencing only plasma conditions in a "thin" surface layer. The requirement of preheating electrons over a fusion reaction length above a threshold-temperature would then involve the third dimension (depth) only through electron heat conduction, which would yield a different numerical value for the energy requirement, albeit the same scaling with density[vii].

## 4. Use of fast ions for fuel compression

The authors of Ref. [1] stress also that "ionic implosion velocities and flow pressures inherent in their scheme are orders of magnitude larger than those achievable with thermal ablation". These high energy densities exist, however, only over extremely small regions around the individual nanostructures, which have line densities $\varrho R$ orders of magnitude too small to produce fusion energy gain! Concerning their possible contribution to global compression these hot ions would make bad use of the invested energy. A direct use of the recoil momentum of these fast ions to accelerate a massive fuel shell would be inefficient, as their high particle velocities and small mass imply a bad use of energy due to the poor rocket efficiency (see, e.g. Ref. [3]). If the role of the ions were to transfer energy deeper into the pellet shell of a large aspect ratio configuration - like indicated in Fig. 5 top - the short duration of efficient laser-nanostructure coupling would limit the acceleration phase, and hence the total coupled energy and ablated mass. A mixing of the regions of hot ions and of residual fuel would cause pre-heating of the target and prevent strong compression. Ultimately the particular geometry of the laser-irradiation, nanostructure, and fast-ion anisotropy indicated in Figs. 1 and 5 appear difficult to reconcile with an efficient spherical implosion.

## 5. Conclusions

Since the seminal paper by Nuckolls et al. [8] strong compression of nuclear fuel has been universally recognized as a central requirement for peaceful applications of inertial fusion [9].

---

[vii] The depth would then be given by $\Delta x \sim \sqrt{\chi_{h.c.} \cdot \tau_{fusion}}$, with both $\chi_{h.c.}$ and $\tau_{fusion} \sim 1/n$.

This requirement for compression in the form of a minimum needed value of $\varrho R$ arises also when fusion burn is to be triggered by fast ions in different forms as proposed in [1] and [2], or even when only energy amplification in a surrounding medium ("convertor") is aimed for. Even for the most favorable D-T reaction prohibitively large required laser energies would be required, if fuel assemblies at solid state densities are considered. Due to the smaller cross-section and higher particle energies, the required $\varrho R$ values for a p-$^{11}$B fuel would even be one order of magnitude, and the single ion energies a factor of six larger than the corresponding D-T values.

Distinguishing technical features of the Marvel proposal are the use of ultra-short intense laser pulses to generate efficiently high energy ions. We show in our comments, that such sub-picosecond systems can be of use for fusion energy production only if the characteristic time scale for fusion reactions can be brought down into this range. This requires, however, a pre-compression of the fuel by some 3 orders of magnitude. Such systems are thus potentially well matched to fast ignitor designs, provided their energy can be delivered to a highly pre-compressed fuel. This appears, however, very difficult to reconcile with nanostructured targets, which certainly could not survive significant compression. The proposals in Ref. in [1] and [2], however, do not address this fundamental difficulty.

Acknowledgements:

The authors thank Dr. Rainer Burhenn for helpful discussions.